# A fuzzy logic based method for efficient retrieval of vague and uncertain spatial expressions in text exploiting the granulation of the spatial event queries

V. R. Kanagavalli
Research Scholar
Sathyabama University
Chennai
kanagavalli.teacher@gmail.com

Dr. K. Raja
Dean (Academics)
Alpha College of Engineering
Chennai
raja_koth@yahoo.co.in

## ABSTRACT

The arrangement of things in n-dimensional space is specified as Spatial. Spatial data consists of values that denote the location and shape of objects and areas on the earth's surface. Spatial information includes facts such as location of features, the relationship of geographic features and measurements of geographic features. The spatial cognition is a primal area of study in various other fields such as Robotics, Psychology, Geosciences, Geography, Political Sciences, Geographic Economy, Environmental, Mining and Petroleum Engineering, Natural Resources, Epidemiology, Demography etc., Any text document which contains physical location specifications such as place names, geographic coordinates, landmarks, country names etc., are supposed to contain the spatial information. The spatial information may also be represented using vague or fuzzy descriptions involving linguistic terms such as near to, far from, to the east of, very close. Given a query involving events, the aim of this ongoing research work is to extract the relevant information from multiple text documents, resolve the uncertainty and vagueness and translate them in to locations in a map. The input to the system would be a text Corpus and a Spatial Query event. The output of the system is a map showing the most possible, disambiguated location of the event queried. The author proposes Fuzzy Logic Techniques for resolving the uncertainty in the spatial expressions.

## 1. INTRODUCTION

Text documents have been the source of information for a long period for many applications. Even though other forms of data are now supported by the commercial databases and web has been the repository of all forms of data, text documents are still a major source of information. A text document with spatial expressions embedded is of interest in various applications. They can be found in documents generated and referred by environment engineers, historians, journalists and tourists. [5] A spatial expression conveys information about objects in space, their locations, or the frame of reference from which objects are viewed. Understanding a spatial expression includes representing the information it conveys so that it can be used for spatial reasoning. Spatial expressions are found in event descriptions where an event is described as a happening, an occurrence. An event is characterized by: who, what, when, where and some form of attribution [9]. A specific branch of information retrieval, geographic information retrieval actually refers to the entire pipeline of extracting geographic referents from text, indexing them into a spatial index, and allowing spatial search of a corpus using the spatial information [21].

This ongoing research work of this author is organized as follows:

- Given a spatial query determine the level of granulation desired by the user from the spatial query.
- Extract the relevant documents matching the spatial query from the text corpus.
  Indexing method used to retrieve the documents is as follows
    o extract spatial expressions containing the query terms from free form text reports
    o Map the spatial expressions extracted into fuzzy functions; in other words model using possibility distribution function.
    o Store the possibility distribution functions using the most optimal representation methods
    o Enable the retrieval of the possibility distribution functions using most optimal indexing methods
    o Implement and test the efficiency of query processing using the aforesaid indexing method.
.

## 2. Related Work

Extracting spatial data from text is researched under various domains. The most popular application areas include Robotics, Industrial Automation, and Text-to-Scene conversion systems. One of the famous text-to-scene conversion systems, WordsEye is based on VigNet, a unified knowledge base and representational system for expressing lexical and real-world knowledge needed to depict scenes from text [23]. Current spatial oriented research works concentrates on the fields of, spatial generalization [12, 13, and 14], spatial normalization [16] and spatial summarization [17] as identified by the authors of the work [11].

Research in the direction of providing natural language directions to the humanoid by voice instruction suggests extracting a sequence of spatial description clauses from the linguistic input; the humanoid then infers the most probable path through the environment given only information about the environmental geometry and detected visible objects [8].

The four functionally distinct tasks for understanding the spatial expression are extracting spatial expressions and mapping spatial expressions onto spatial primitives. The authors of [5] explore the notion of spatial expression understanding relative to the domain of newspaper captions. Detecting and extracting archaeological events described in natural language text are experimented in the literature [6].

Qualitative spatial information can be represented by associating qualitative relations with fuzzy sets. Starting with the concept of absolute distance, they have extended the metric notion of proximity to non-metric notions of proximity [7].

The issues of vagueness that arise in the spatial language that people use when making queries (near to, within walking distance, etc.) and the vagueness in the spatial extent of some geographical regions such as neighborhoods within cities are handled by a body of literature both in GIS and Natural language processing. The fuzzy methods are employed to model interpretations of spatial language prepositions and of the extent of vague places. In both cases, the parameters for these models are based on data obtained from Web pages that include instances of the various types of vague language and vague neighborhoods. Thus, websites that describe hotels often include textual descriptions that describe their location using phrases such as 'within walking distance' of some prominent location such as a well-known local landmark. The spatial language expressions can then be compared with knowledge of the actual distances involved. The set of actual distances that correspond with the use of a particular expression can be used to create a fuzzy-set membership function.

A query can be treated as a generation process [10]. Given a sequence of terms in a query the probabilities of generating these terms according to each document model can be computed. The multiplication of these probabilities is then used to rank the retrieved documents: the higher the generation probabilities, the more relevant the documents to the given query.

The granulation level of the query can be used to choose the best suited index [11]. For complex queries composed of different grained spatial features a default well suited index is used. A classical textual tokenization preprocessing is followed by location of spatial named entities using typographic and lexical rules. The next step is to mark the absolute spatial features and relative spatial features. The gazetteers are used to validate the absolute spatial features. The first level of spatial index is built from the validated spatial features which are later refined to second level spatial index supporting information retrieval capabilities.

Other techniques such as divide and conquer, integration and dynamic visualization are also used for searching spatial and non-spatial information [22]. The search outcome integrates both documents and map data providing easy comparison and verification of the information extracted from different sources.

Ranking functions computed from the spatial terms, geographic footprints of the query and the documents are used for the retrieval of the most relevant documents given a query. A detailed analysis of text-first retrieval and map-first retrieval is also discussed in the literature [25].

## 3. Issues in handling Vague Spatial expressions and Queries

Indirect spatial reference is any way to describe a location without using coordinates, generally using a geographic feature to uniquely identify a place. They are important because they are a very common means by which observations of other attribute information are tied to a place. They can serve as a means to link the attribute data to coordinate descriptions of the place to which the attribute data apply.

Spatial referencing is a qualitative process for humans whereas it is a quantitative process for computers. For effective spatial query processing, there is a need of a mechanism to convert the qualitative data to quantitative data. The issues associated with the extraction and representation of vague event descriptions from web documents is discussed in literature [3].

- Vague Spatial Regions and Vague Spatial Relations result in indeterminate boundaries in GIS Modeling.
- There is little support in digital gazetteers for imprecise, vague spatial references.
- Generally digital Gazetteers refer to a point inside the geographic feature (usually the centroid) and do not refer to the actual spatial extent of the geographic features.
- There is currently very less work to incorporate the network features.
- To collect information from multiple documents to resolve the uncertainty in spatial location

Spatial queries are often found in geographic information systems and there is still lot of issues in handling spatial expressions [1].

- Detecting the geographical information within users queries and text documents
- Disambiguating the place names to find the intended **one**
- Interpreting the geometric location of vague place names
- Spatially and thematically indexing the text documents within a GIR System
- Information retrieval model to pick up the relevant documents out of the library and ranking the degree of relevance according to their spatial and non-spatial properties.
- Effective user interface
- Approaches to evaluate the success of a GIR System

## 4. Spatial information in Text

As there is significantly more textual than spatial data in current geo search engines, it is important to focus on the textual aspect of the problem. Mining spatial information from text reports involves natural language processing and text processing wherein the spatial components expressed in the form of geocodes, place names, addresses etc., would be read in by the system.

The question that arises in this discussion is, whether it is feasible to use the current techniques of modeling the spatial information embedded in text when maps are so often used in many applications? The authors reiterate and go by the argument that even though the maps are used in many applications still there is lot of scope in modeling the spatial information from the text since there is lot of text documents available in text databases in the form of reports, documents, scientific descriptions, tourist reports etc.,

Modeling spatial information involves mapping the spatial descriptions on to the logical space using mathematical functions. So far, the spatial descriptions are mapped using the probability density functions as they possess the properties of formality, practicality, generality and effectiveness. Existing solutions that employ the probability theory are known to be effective and scalable.

The authors of this work would like to argue on the same grounds the feasibility of using fuzzy logic techniques for modeling the spatial information extracted from the text reports.

## 5. Handling Uncertain spatial expressions in Text

Uncertainty can be categorized into two types, vagueness and ambiguity. Vagueness is associated with the lack of clarity of the definition of a class to which a given element belongs whereas ambiguity is associated with the lack of clarity in information about the given element and there exists clear information about the class to it belongs. In the context of spatial information embedded in text reports, there exists a scope of both vagueness and ambiguity. Vagueness exists whenever there is a spatial location specified but it cannot be inferred to which sense it is specified in the document and ambiguity arises whenever there is insufficient information about a spatial location specified in the document. Modeling spatial information is a vital process to visualize the spatial components in the text document. The quality of a model is evaluated based on the complexity of the model and credibility of the model. Uncertainty is a vital factor in increasing the credibility of the model by reducing the inherent complexity present in the real world.

The spatial location names or the spatial expressions associated may be vague which makes it difficult to search the desired location using geographic gazetteers since the gazetteers employ official names of the spatial locations which necessitate the fuzzy extent of the spatial location to be models. Being able to interpret such terms will help in analyzing the geographic context of documents and in interpreting user queries that employ vague spatial language [4].

Fuzzy logic is based on the theory of fuzzy sets, a theory which relates to classes of objects with un-sharp boundaries in which membership is a matter of degree. Fuzzy logic allows computing with words closer to human intuition rather than numbers thus allowing the tolerance for imprecision [31]. Fuzzy logic is very apt for systems that would comprehend spatial information from the text repositories since the task calls for more of common-sense and human cognition than any other mathematical methodology. The capability of fuzzy sets to express gradual transitions from membership to non-membership and vice versa provides us with a meaningful representation of vague concepts expressed in natural language. Use of fuzzy logic bridges the gap that exists between the real world imprecise or ambiguous system and the perfect objects built in a model. Fuzzy spatial reasoning is a method for handling different types of uncertainty inherent in almost all spatial data. Qualitative spatial information can be best represented using fuzzy logic.

## 6. Proposed Work

The spatial query from the user is fed into the system. The proposed system retrieves relevant documents from the text corpus. The top *n* documents (based on the ranking) are retrieved from the text corpus. The disambiguation may be associated with the spatial query or with the event description in the text document.

Algorithms used for retrieving the relevant documents from the corpus so far have used the footprint of the document, toe print of the document, etc., The algorithms used so far can be technically be divided into two classes either geo-first algorithms or text-first algorithms. The relevance of a document to the given query can be determined by the number of occurrences of the place name in the documents, the place of occurrence, the occurrence of the related places, spatial descriptors etc., This work differs from the existing body of literature in the sense that the spatial relevance and importance of the documents, based on the granulation level in the query are determined using fuzzy distribution functions

The proposed system disambiguates the query and resolves the uncertainty in the text document by considering the fuzzy, ambiguous event description in multiple documents. The advantage of the vector space model is that it allows spatial information to be handled the same way as thematic information with the proper use of proper ontology of places [19, 20].

*Taxonomic knowledge* of task-relevant geographic layers should be taken into account to obtain descriptions at different granularity levels.

Step 1: Accept the user's spatial keyword query involving an event.
Step 2: Determine the level of granulation required by the user by analyzing the spatial query entered.
Step 3: Identify the most certain areas and least certain areas of the related event
Step 4: Retrieve the documents from the text corpus that most possibly contains the event descriptions and the spatial references of most certain areas.
Step 5: Extract identifiable spatial references, vague and imprecise spatial references using text engineering tool.

Step 6: Use the granulation level to decide the spatial data type for representing the spatial attribute and to determine the possibility of the document matching the user query requirements

Step 7: Use fuzzy logic techniques to model the spatial references.

Step 8: Incorporate the results in a GIS

Step 9: Display the resolved spatial references

## 6.1 Fuzzy Vector Space Retrieval Method

Combining fuzzy logic techniques with existing vector based model ensure the simplicity and formalism of the logic based model, and the flexibility and performance of the vector model. Unlike the Boolean model that is based on binary decision criterion {relevant, not relevant}, fuzzy logic expresses relevance as degrees of memberships (e.g., document | query could have a relevance measure with the following degrees of membership: 0.7 highly relevant and 0.5 moderately relevant and 0.1 not relevant) [30]. Fuzzy reasoning is used to determine the importance of a word thus reducing the dimensionality of the vector space and the retrieval result is grouped automatically according to page contents using the vector space model method, the frequency of word appearance and the fuzzy reasoning.

The vector space model is slightly modified where each document is modified as a vector v in the t dimensional space R and the spatial term frequency is defined as the number of occurrences of term spatial term S in the document D, that is, $S_F(D; S)$. The difference between calculating the ranking based on the non-spatial and spatial term is the identification and disambiguation of spatial terms. The (weighted) spatial term-frequency matrix $S_{WF}(D; S)$ measures the association of a term S with respect to the given document D.

Using the traditional vector space method the weight of the spatial term is computed using the following formula,

$$S_{WF} = SF \times IDF = SF \times (\log_2(N/n)+1) \quad (1)$$

where $IDF$ is the Inverse Document Frequency, N is the total number of documents and n is the number of documents involving the spatial term S. Once the importance of the given spatial term in the document is identified, the next step is to find the spatial similarity between two different documents present in the text corpus. The spatial similarity $S_s(D_i, D_j)$ between two documents $D_i$ and $D_j$ is defined as the measure of possibility that both the documents are referring to the same geographic locations.

By following the traditional information retrieval formula, the spatial similarity of the document is measured as the cosine of the angle between the two document vectors which is calculated as follows:

$$\sum_{k=1}^{m}(S_{WFi} \times S_{WFj})/[(\sqrt{\sum_{k=1}^{m}(S_{WFi})^2} (\sum_{k=1}^{m}(S_{WFj})^2]$$

Where m is the total number of documents, and $S_{WFi}$ and $S_{WFj}$ are the weighted frequency of the spatial term in the document $D_i$ and $D_j$. The authors propose to apply intuitive fuzzy logic techniques for the retrieval of the documents. It is proposed to be better than the existing methods since, unlike the crisp logic which decides whether the document is relevant or not, fuzzy logic allows computing the relevance as degrees of memberships.

Though there are works using fuzzy logic techniques for the information retrieval, there is no fuzzy logic function applied for the ranking of the documents retrieved. The proposed method would find out whether the terms are spatial references by referring to gazetteer and also applying fuzzy logic memberships and using fuzzy rule base. The word sense disambiguation techniques are used to find out whether a spatial resembling term does actually refer to a geographic location or not. The fuzzy membership functions are assigned to the terms with the use of the fuzzy rule base and the weight of the spatial referring term is calculated based on that. This is much different from the traditional document classification methods.

The fuzzy rules are generated using the fuzzy membership functions for the documents, using the $S_{WF}$ calculated using the formula (1). If SWF is high then the document is highly relevant to the current query involving the spatial term S, if SWF is low, then the document's relevance is estimated to be low. This graded relevance would help user to resolve the level of uncertainty associated with the documents and the relevance of the document with respect to the query posed by the user to the system.

.

## 6.2 Handling Granularity

Granularity is the concept of breaking down an event into smaller parts or granules such that each individual granule plays a part in the higher level event. Semantic granularity addresses the different levels of specification of an entity in the real world, while spatial granularity deals with the different levels of spatial resolution or representation at different scales [24]. The granularity of the spatial information or spatial query can be used to understand the user's need and provide much relevant information to the user. It can be exploited in multiple domains such as business sectors, disaster response systems, environmental engineering and city planning. The early work of this author throws light on exploiting the granulation of the spatial information for increasing the effectiveness of business activity monitoring [27].

A granularity structure exists only if at least 2 levels of information are present in text, such that the events in the coarse granularity can be decomposed into the events in the fine granularity and the events in the fine granularity combine together to form at least one segment of the event in the coarse granularity [28, 29]. The authors of the above works have presented the figure [2] in which $G_c$ represents the phrase or sentence with coarse granularity information and $G_f$ represents a phrase or sentence with fine granularity information. Three possible links connect the objects of coarse granularity and the objects of fine granularity - part-whole relations between events, part-whole relations between entities, and a causal relation between the events in the fine granularity and the events in the coarse granularity. The literature study reveals that there are six types of part-whole relationships, Component-Integral, Member-Collection, Portion-Mass, Stuff-Object, Feature-Activity, Place-Area. This work concentrates on the part-whole relationships including place-area whereas the existing works concentrates on the temporal granulation, and the feature-activity. The difference between afore said work and this work is that the authors of this work are more concerned with the granularity between the spatial entities whereas the referenced work concentrates on the causal relation between the events.

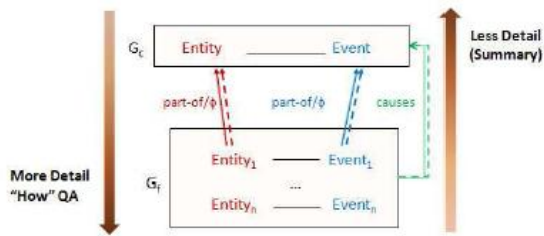

**Figure 1: Granularity in Natural Language Descriptions**

## 7. Conclusion

The authors are retrieving the documents from the text corpus which are relevant to the spatial queries and are different from the traditional Boolean ranking since the documents are ranked on the basis of relevance using fuzzy logic techniques. The fuzzy membership functions determine the spatial relevance of the documents and the fuzzy rules decide the relevance. The spatial similarity between two documents is also evaluated on basis of the fuzzy rule base. The granularity of the query and the spatial information present in the text are used to resolve the uncertainty of the spatial information. Possibility functions, Fuzzy logic techniques are used to model the uncertainty of the spatial information present in the text instead of the probability logic.

The limitations of this work is that it would answer spatial queries involving spatial attributes only and not spatial queries involving geometric shapes since it is querying the textual data and not the spatial database.